# From one space dimension to two space dimensions in special relativity

Bernhard Rothenstein, "Politehnica" University of Timisoara, Dept. of Physics, Timisoara Romania
bernhard_rothenstein@yahoo.com

## 1. Introduction

The derivation of the Lorentz-Einstein transformations (LET) for the space-time coordinates of the same event starts in many textbooks devoted to the subject starts with the derivation of the transformation for the time coordinates

$$t = \frac{t' + \frac{v}{c^2}x'}{\sqrt{1 - \frac{v^2}{c^2}}} \qquad (1)$$

and for the x(x') space coordinate

$$x = \frac{x' + vt'}{\sqrt{1 - \frac{v^2}{c^2}}} \ . \qquad (2)$$

We measure the x(x') space coordinates along the overlapped OX(O'X') axes of the K(XOY) and K'(X'O'Y') inertial reference frames in the well known standard arrangement.
In order to extend the results to two space dimensions Authors [1],[2],[3],[4] add to them

$$y = y' \qquad (3)$$

considering that distances measured perpendicular to the direction of relative motion are relativistic invariants because of thought experiments and of the principle of relativity. Sophisticated students often ask if the transition from one-space dimension to two space dimensions does not affect relation (1) which depends only on the x coordinate.
Mathews Jr. [5] presents relativistic velocity transformation without using the (LET). We use them in order to present a direct two-space dimensions approach to the (LET). The purpose of our paper is to prove that the transformation for the time coordinate (1) depends only on the x space coordinate measured along the overlapped axes.

## 2. A direct two-space dimensions approach

Mathews Jr. [5] presents relativistic velocity transformation without using the (LET). We use them in order to present a direct two-space dimensions approach to the (LET).
Consider a particle that goes at the common origin of time (t=t'=0) through the origins O and O' located at that very moment at the same point in space. It moves with velocity **u**($u_x, u_y$) relative to K, but with velocity **u'**$(u'_x, u'_y)$ relative to K'. The motion takes place along a direction θ when detected from K but along a direction θ' when detected from K'. We measure the angles θ and θ' from the positive directions of the overlapped axes OX(O'X'). The relativistic transformation equations are [5]

$$u_x = u' \frac{\cos\theta' + \frac{v}{u'}}{1 + \frac{vu'}{c^2}\cos\theta'} \qquad (4)$$

$$u_y = u' \frac{\sqrt{1 - \frac{v^2}{c^2}} \sin\theta'}{1 + \frac{vu'}{c^2}\cos\theta'} \qquad (5)$$

$$u = u' \frac{\sqrt{(\cos\theta' + \frac{v}{u'})^2 + (1 - \frac{v^2}{c^2})\sin^2\theta'}}{1 + \frac{vu'}{c^2}\cos\theta'} \qquad (6)$$

Relation (6) leads to the following relativistic identity

$$\frac{1}{\sqrt{1 - \frac{u^2}{c^2}}} = \frac{1 + \frac{vu'}{c^2}\cos\theta'}{\sqrt{1 - \frac{v^2}{c^2}}\sqrt{1 - \frac{u'^2}{c^2}}} \qquad (7)$$

which remains an identity even if we multiply both its sides with a relativistic invariant, say a proper time interval $d\tau$.

In accordance with the scenario we follow, at a given time, we find at a given point of the plane defined by the axes of the two involved reference frames, a clock C(x,y) at rest in K, a clock C'(x',y'). At the same time and at the same point we find a clock $C^0$ that moves with velocity **u** relative to K and with velocity **u'** relative to K'. Let dt, dt' and dτ be infinitesimal changes in the readings of the three clock respectively. In accordance with the formula which describes the time dilation effect, that can be derived without using the (LET) [6] we should have

$$dt = \frac{d\tau}{\sqrt{1 - \frac{u^2}{c^2}}} \qquad (8)$$

$$dt' = \frac{d\tau}{\sqrt{1 - \frac{u'^2}{c^2}}} \qquad (9)$$

Multiplying both sides of relation (7) with dτ and taking into account relations (8) and (9) it becomes

$$dt = dt' \frac{1 + \frac{vu'_x}{c^2}}{\sqrt{1 - \frac{v^2}{c^2}}} = \frac{dt' + \frac{v}{c^2}dx'}{\sqrt{1 - \frac{v^2}{c^2}}} \quad . \tag{10}$$

depending only on the x' coordinate. Starting with the relativistic identity

$$\frac{u_x}{\sqrt{1 - \frac{u^2}{c^2}}} = \frac{u'_x + v}{\sqrt{1 - \frac{v^2}{c^2}}\sqrt{1 - \frac{u'^2}{c^2}}} \tag{11}$$

and multiplying both its sides with dτ we obtain

$$dx = \frac{dx' + vdt'}{\sqrt{1 - \frac{v^2}{c^2}}} \quad . \tag{12}$$

Starting with the relativistic identity

$$\frac{u_y}{\sqrt{1 - \frac{u^2}{c^2}}} = \frac{u'_y}{\sqrt{1 - \frac{u'^2_y}{c^2}}} \tag{13}$$

we obtain multiplying both its sides with dτ

$$dy = dy' \tag{14}$$

distance measured perpendicular to the direction of relative motion being a relativistic invariant.

**3. Conclusions**

Mathews Jr. [5] derives relations (5) and (6) based on thought experiments, time dilation, and length contraction and kinematical considerations. Based on his results we can consider that relation (10) has its roots in the same facts showing that transition from one-space dimension to two-space dimensions does not affect the transformation of the time coordinate.